# Thermodynamics of hydrogen vacancies in MgH$_2$ from first-principles calculations and grand-canonical statistical mechanics

R. Grau-Crespo,[1,*] K. C. Smith,[2] T. S. Fisher,[2] N. H. de Leeuw,[1] and U. V. Waghmare[3]

[1] *Department of Chemistry, University College London, 20 Gordon Street, London WC1H 0AJ, UK. Corresponding author's email: r.grau-crespo@ucl.ac.uk*

[2] *School of Mechanical Engineering and Birck Nanotechnology Center, Purdue University, West Lafayette, IN 47907, USA*

[3] *Theoretical Sciences Unit, Jawaharlal Nehru Centre for Advanced Scientific Research, Jakkur Campus, Bangalore 560 064, India.*

Ab initio calculations and statistical mechanics are combined to elucidate the thermodynamics of H vacancies in MgH$_2$. A general method based on a grand-canonical ensemble of defect configurations is introduced to model the exchange of hydrogen between crystalline MgH$_2$ and gas-phase H$_2$. We find that, at temperatures and hydrogen partial pressures of practical interest, MgH$_2$ is capable of accommodating only very small concentrations of hydrogen vacancies, which consist mainly of isolated defects rather than vacancy clusters, contrary to what is expected from a simple energetic analysis.

PACS numbers: 61.72.Bb, 61.72.jd, 61.50.Ah, 61.50.Ks

## I. INTRODUCTION

The storage of hydrogen at high densities in an accessible state and at reasonable cost represents a major technological challenge and is perhaps the primary barrier to the use of hydrogen as a clean energy carrier for transportation [1]. Magnesium hydride (MgH$_2$) is the parent compound of an important group of hydrogen storage materials and exhibits high volumetric and gravimetric storage density (7.7% and 55 kg/m$^3$, respectively). Although the practical potential of pure MgH$_2$ is limited by its high thermodynamic stability and slow kinetics, mechanical treatment and chemical alloying have been shown to improve the kinetics and to reduce the desorption temperatures [2].

The release and loading of hydrogen in metal-H systems generally occurs via the transformation between two phases: an α phase with hydrogen in lower concentrations occupying interstitial spaces in the metallic lattice (*e.g.* hcp Mg), and a β phase, which is an ionic hydride crystal (*e.g.* rutile structure of MgH$_2$) with hydrogen vacancies in low concentration. The mechanisms of these transformations should be understood at a fundamental level to achieve a rational design of hydride materials for hydrogen storage. Sophisticated models have therefore been proposed to describe the thermodynamics of hydride formation and decomposition [3-5], and *ab initio* electronic structure calculations are being increasingly used to provide microscopic information about these processes [6-10].

We report here the results of a theoretical investigation of the microscopic configurations and related thermodynamics of hydrogen vacancies in the magnesium hydride β phase, near the transformation point to the α phase. The properties of hydrogen vacancies in this hydride have been investigated previously using density functional theory (DFT), and it was concluded that vacancy pairs tend to form more readily than isolated vacancies at the surface of MgH$_2$ [11]. Du et al. also suggest that hydrogen vacancy diffusion is unlikely to be rate-determining because of low activation energy (< 0.7 eV) compared to surface desorption. However, a more complete analysis of this issue should take into account the concentration and nature of these vacancies, which depend on the temperature and partial pressure of hydrogen. Thus, this work is a first attempt to incorporate finite temperature and pressure effects in modeling the properties of hydrogen vacancies in hydrides, accounting for both the solid-gas equilibrium and the association and disorder of defects in the solid. In order to achieve this task, we also introduce a general methodology for the investigation of the distribution of ions in a solid as a function of the pressure of a gas phase of the same species.

## II. CALCULATION DETAILS

DFT calculations were performed to obtain the energies and relaxed geometries for different defect distributions in the 2x2x2 and 2x2x3 supercells of the rutile-like MgH$_2$ structure (Fig. 1). We employed a generalized gradient approximation (GGA) functional built from the Perdew and Zunger [12] local functional, with the spin interpolation formula of Vosko *et al.* [13] and the gradient corrections by Perdew *et al.* [14]. A cutoff energy of 500 eV for the plane wave basis set and a mesh of 2x2x3 (or 2x2x2 for the larger supercell) $k$-points for Brillouin zone integrations were employed. The interaction between the valence electrons and the core was described with the projected augmented wave (PAW) method [15], as implemented in VASP by Kresse and Joubert [16]. For Mg, core orbitals up to $2p$ were kept frozen to the reference levels. The ionic positions were relaxed until the forces on each atom were less than 0.01 eV/Å. The relaxed parameters for the non-defective unit cell were $a$ = 4.4924 Å, $c$ = 3.0049 Å, while the position parameter for the H ions was $u$ = 0.3046 (experimental values are 4.5168 Å, 3.0205 Å, and 0.3060, respectively [17]).

For each composition, characterized by $n$ vacancies, there are $N!/n!(N-n)!$ ionic configurations, where $N$ = 32 is the number of hydrogen sites in the 2x2x2 supercell. However, only the symmetrically inequivalent configurations were calculated, as determined using the SOD program [18], resulting in 1, 1, 17 and 32 distinct configurations for $n$= 0, 1, 2 and 3, respectively. A few configurations were also calculated in a 2x2x3 supercell, as explained below, in order to check the convergence of our results with the supercell size. For each configuration, all different spin solutions were calculated in order to determine the ground state. The vacancy formation energies (per vacancy) were calculated as:

$$VFE = \frac{1}{n}\left(E[\text{Mg}_{16}\text{H}_{32-n}] + \frac{n}{2}E[\text{H}_2] - E[\text{Mg}_{16}\text{H}_{32}]\right) \quad (1)$$

where $E[\text{Mg}_{16}\text{H}_{32}]$ and $E[\text{Mg}_{16}\text{H}_{32-n}]$ are the calculated energies of the pure and defective supercells, respectively, and $E[\text{H}_2]$ is the energy of an isolated hydrogen molecule, calculated in box of 10 Å × 10 Å × 10Å. For the finite-temperature analysis, vibrational contributions to the free energies of the configurations were estimated in the harmonic approximation:

$$\Delta F_{vib} = k_B T \sum_i \ln(2\sinh(\frac{\hbar\omega_i}{2k_B T})) \quad (2)$$

from the Γ-point phonon frequencies $\{\omega_i\}$, which were calculated from the energy-minimized structures (no free energy minimization was performed).

## III. RESULTS AND DISCUSSION

### A. Configurational spectra and vacancy formation energies

The calculated DFT energies (Fig. 2) show significant variations in stability among the different vacancy distributions for each composition. A simple inspection of the results reveals that the most stable configurations with multiple vacancies per cell involve the formation of vacancy clusters. In the case of $n=2$, for example, the lowest-energy configuration consists of two vacancies located at the shared edge of one $MgH_6$ octahedron, corresponding to the shortest H – H distance in the perfect structure (2.48 Å). The second most stable configuration also has two vacancies in the same octahedron but with a different orientation of the pair, in this case involving the shared corner of the octahedron corresponding to the next shortest H – H distance in the structure (2.75 Å). The other configurations with composition $n=2$ are much higher in energy. For $n=3$, a smaller energy gap exists between the configurations, but again the lowest energy state exhibits vacancy aggregation: two vacancies in the shared edge and one in the shared corner of the same octahedron. The VFE, as obtained from Eq. (1), is always lower for a vacancy in a cluster (1.04 and 1.07 eV for the most stable di-vacancy and tri-vacancy configurations, respectively) than for an isolated vacancy (1.41 eV).

We have verified that these results are not an artifact resulting from the interaction of the defects with their images in the periodic supercell. The 2×2×2 supercell employed in our calculations has dimensions 8.98 Å × 8.98 Å × 6.01 Å, and therefore the most significant interaction of the defects with their images will occur along the *c* axis. We have therefore calculated the VFEs of different species in a 2×2×3 supercell (8.98 Å × 8.98 Å × 9.01 Å), and we have found that they change (increase) by only 1 meV, 8 meV and 43 meV for the most stable single vacancy, di-vacancy and tri-vacancy configuration, respectively, with respect to the 2×2×2 supercell. Obviously, the interaction with the images increases with the size of the defects, but even for the largest defect the effect remains small. The energetic preference for vacancy aggregation is therefore clear from our results and consistent with earlier work [11].

Fig. 3 shows the charge density distributions around mono-vacancies and di-vacancies, and their corresponding electronic densities of states (DOS). In both cases, the electrons left behind by the removed hydrogen remain localized in the defect region, as expected given that the species present in the solid cannot be reduced easily [19-21]. In the case of $n = 1$, the ground state is spin-polarized with one unpaired electron localized on the vacancy site, and a DOS analysis reveals that the defect is responsible for an occupied spin-up state and an unoccupied spin-down state localized in the band gap (at ~4 eV in the non-defective solid), with an exchange splitting of ~1 eV. In the case of $n = 2$, the ground state is not spin-polarized, as the two electrons localized at the vacancy sites prefer to have opposite spin orientations. The defect gap levels in the DOS are now symmetric in spin, with a larger energy separation between the unoccupied and the occupied states, with the former closer to the top of the valence band and the latter closer to the bottom of the conduction band of $MgH_2$ compared to the mono-vacancy. This rearrangement of energy levels resembles the formation of a bond and is responsible for the stabilization of di-vacancies with respect to isolated mono-vacancies.

## B. Grand-canonical statistical mechanics analysis

The energetic preference for vacancy aggregation does not necessarily dictate that vacancies will predominantly form clusters in the solid. In particular, if the overall concentration of vacancies in the material is very low, the number of ways in which two "isolated" vacancies are accommodated is much higher than the number of di-vacancy configurations in the corresponding very large supercell. In other words, the probability of occurrence of mono-vacancies could be higher than the probability of di-vacancies (or of higher-order clusters), regardless of the relative formation energies, provided that the vacancy concentration is low enough. In principle, we can represent this situation in a typical canonical formulation at a fixed concentration (as used, for example, in Refs [22-25]), where the probability of a particular configuration (*m*) depends on both its energy $E_m$ and its degeneracy $\Omega_m$ as:

$$P_m = \frac{\Omega_m}{Z} \exp\left(-\frac{E_m}{k_B T}\right) \qquad (3)$$

where *m* is an index over all the inequivalent configurations, $k_B$ is Boltzmann's constant and $Z$ is the canonical partition function. For low concentrations, large supercells are required, and the cumulative degeneracy of all configurations where the vacancies are isolated will be much higher than the degeneracy of any cluster configuration. However, it is clear that this canonical representation is impractical in the low concentration limit, as it requires supercells that are too large to be treated at the DFT level. Furthermore, in this canonical representation we must know the vacancy concentration *a priori*, while it would be more useful to be able to predict this concentration based on DFT results.

We therefore introduce a grand-canonical formulation that enables the calculation of vacancy concentration and distribution in a solid at equilibrium with a gas phase of the vacant species (hydrogen in this case), using inputs from DFT calculations in smaller supercells. The defective solid with any given concentration of vacancies is represented as a grand-canonical ensemble of configurations, including cells with stoichiometric composition (*n*=0) and with defects (*n*>0) in the relevant proportion. The probability of occurrence at temperature *T* of any particular configuration is given by:

$$P_{nm} = \frac{\Omega_{nm}}{\Xi} \exp\left(-\frac{E_{nm} + n\mu}{k_B T}\right) \qquad (4)$$

where now $E_{nm}$ and $\Omega_{nm}$ are the energy and degeneracy of the $m^{th}$ inequivalent configuration with *n* vacancies, $\mu$ is the chemical potential of hydrogen in the structure, and $\Xi$ is the grand-canonical partition function. In order to introduce the effect of the gas pressure, the chemical potential of hydrogen in the hydride is assumed to be identical to the potential per atom in the gas phase (solid – gas equilibrium condition):

$$\mu = \frac{1}{2} g_{H_2}(T, p_{H_2}) = \frac{1}{2}\left(E[H_2] + E_{ZP}[H_2] + \Delta g_{H_2}(T, p_{H_2})\right) , \qquad (5)$$

where $E_{ZP}[H_2]$= 0.270 eV is the zero-point energy of the molecule, calculated from the theoretical vibrational frequency $\omega$=4357 cm$^{-1}$ (experimentally, $\omega$=4401 cm$^{-1}$ [26], which corresponds to $E_{ZP}[H_2]$= 0.273 eV). The term $\Delta g_{H_2}$ represents the change of the free energy per gas molecule from 0 K to temperature *T* in a gas at pressure $p_{H_2}$. The values of this pressure- and temperature-dependent term are difficult to obtain from first-principles, and theoretical estimates are inaccurate because of the poor description of intermolecular dispersion interactions within DFT. Therefore, ab initio studies of solid-gas equilibrium thermodynamics commonly employ a combination of experimental information and/or ideal gas expressions to approximate this contribution [27-29], and in the present case we have obtained this information directly from thermodynamic tables [30]. We note that, in order to be consistent, the molecular energy contribution to the hydrogen gas chemical potential was calculated with the same DFT method used for the hydride. The molar fraction $\delta$ of vacancies at any temperature and hydrogen partial pressure is then obtained as:

$$\delta = \frac{1}{N} \sum_n n \sum_m P_{nm}. \qquad (6)$$

Using these expressions, we can represent a defective solid with vacancy concentrations ($\delta$) much smaller than the concentrations ($n/N$) explicitly calculated in the defective supercells, without requiring a larger cell. We can visualize this case as a collection of a large number of non-defective cells together with a few defective ones (high and low $P_{nm}$ probabilities for $n=0$ and $n>0$, respectively). This approach is valid as long as the vacancy formation energies for the most stable species are well converged with respect to cell size, which we have confirmed for our study, as reported above.

The calculations for the configurational equilibrium of vacancies in MgH$_2$ were first performed excluding vibrational contributions. All configurations for $n = 0, 1, 2$ and $3$ were included in the analysis, and the probabilities from Eq. (4) were evaluated using the DFT energies. In a second step, apart from the $n = 0$ and $1$ configurations, only the most stable, cluster-forming configurations were considered for $n > 1$ (two for $n = 2$ and three for $n = 3$, as shown in Fig. 1), and the probabilities in this case were evaluated by adding the vibrational free energy contribution to the energy of each configuration. The vibrational contribution at each temperature was obtained from the phonon frequencies calculated from the structure optimized at 0 K, *i.e.*, no explicit free-energy minimization was performed. We have checked that the ensemble truncation does not affect the predicted concentrations of each species, because at low vacancy concentrations the contribution of configurations with $n \geq 1$ to the partition function is negligible. For example, if we take the energy of the non-defective solid as a reference, the contribution of the $n=0$ configuration ($E_{01}=0$) to the grand-canonical partition function will be 1, while the next largest contribution comes from the configuration with $n=1$ and is only of $\sim 10^{-8}$ at $T=600$ K and $p_{H2}=3.8$ bar, which is the experimental α-β transition pressure at that temperature. Therefore, in order to obtain the effect of vibrational contributions on the relative occurrence of one configuration, we need only to obtain the vibrational free energy of that particular configuration (apart, of course, from the free energy of the $n=0$ configuration). The discussion below thus refers to results for a few important configurations, including vibrational effects. Pressure effects were included only via the pressure dependence of the chemical potential in Eq. (5), as zero external pressure was assumed in all DFT calculations. The direct effect of high pressures on the crystal structure of the hydride would be considerable, including a phase transformation at ~4 kbar [31], but our approximation is adequate for studying the thermodynamics of hydrogen vacancies at the much lower pressures considered here.

The pressure-composition isotherms at 600 – 800 K are shown in Fig. 4. The horizontal portions (labeled α+β) of each isotherm correspond to an equilibrium mixture of α and β phases. The values of equilibrium pressure at each temperature were taken from experimental data [32]. The β part of each isotherm was calculated from DFT results using the method described above. Higher concentrations of vacancies are predicted for lower hydrogen partial pressures and higher temperatures, as expected. The intersection of this β line with the horizontal part of the isotherm determines the maximum fraction of vacancies in the hydride at that particular temperature, and this fraction is predicted to be very small, varying between $10^{-8}$ and $10^{-6}$ for temperatures in the range 600 - 800 K. This result agrees with experimental observations by Stampfer *et al.* [32], who found that the hydride in equilibrium with the α phase has an H/Mg ratio of 1.99 ± 0.01 over the investigated range of temperatures (713 – 833 K). On the other hand, Belkbir *et al.* [33] reported significant departures from the stoichiometric H/Mg = 2 composition, with a clear temperature dependence between 613 and 648 K—a behavior that is not supported by our calculations. As noted by others [34], the isotherms of Belkbir *et al.* are not horizontal in the α+β region, indicating that their samples probably had not reached equilibrium.

The β→α transformation provides the mechanism for release of hydrogen gas, and therefore the distribution of vacancies in the hydride under α-β equilibrium conditions is expected to affect the diffusion of hydrogen inside the material in practical applications [11]. The predicted concentrations of each vacancy species at the transition point (Table 1) clearly reveal that the relative abundance of the different species is not well correlated with the VFEs. Di-vacancies, for example, although energetically favorable with respect to isolated vacancies, are less abundant under these conditions by one or two orders of magnitude. Tri-vacancy clusters are even rarer, despite the fact that their formation energies are relatively low. In the context of a canonical formulation, with a fixed number of vacancies in a much larger supercell, this effect would arise from the much higher degeneracy of configurations with isolated vacancies compared to those with clusters. Equivalently, in our grand-canonical treatment, with much smaller cells, this is a consequence of introducing the $H_2$ chemical potential in the energy balance: weakly negative values of $\mu$ stabilize configurations with smaller values of $n$, as seen in Fig. 5. Inclusion of vibrational effects leads to a significant increase in the predicted vacancy concentrations (*e.g.*, by a factor of ~25 at the transition pressure for 600 K, while the formation energy of isolated vacancies is lowered by 0.16 eV at this temperature), but they do not strongly alter the relative abundance of different defect species.

The dominance of mono-vacancies extends over most pressures and temperatures of interest, as reflected in the linearity of the isotherms on the logarithmic plot of Fig. 4. A simple check reveals that neglecting the multi-vacancy terms in Eq. 4 and using the ideal gas expression for $\mu$ in Eq. 5 yields a dependence of the form $\delta \sim (p_{H_2})^{-1/2}$. Some departure from this behavior, associated with the occurrence of vacancy clusters, can only be noticed if the β zones of the isotherms are extrapolated below the transformation pressure (dashed lines in Fig. 4). On the other hand, the small deviation from the linear shape of the isotherm at high pressures is related not to vacancy association but rather to the non-ideality of the hydrogen gas, which was taken into account in the chemical potentials.

## IV. CONCLUSIONS

Our combination of DFT calculations with grand-canonical statistical mechanics has made possible the prediction of defect composition isotherms in $MgH_2$ as a function of hydrogen gas pressure. We conclude that, although vacancy aggregation is energetically favorable due to the spin pairing of defect electrons, most hydrogen vacancies in $MgH_2$ are isolated at temperatures and pressures of practical interest for hydrogen storage. Under these conditions, the concentrations of vacancies are predicted to be small enough to keep a stoichiometric (H/Mg=2) composition of the hydride within the precision limits of typical measurements. Our results therefore suggest that the slow kinetics of hydrogen diffusion in pure magnesium hydride, which is one of the factors limiting its practical applications in hydrogen storage, is related to the low concentration of vacancies at the conditions of hydrogen release.

## ACKNOWLEDGMENTS

Via our membership of the UK's HPC Materials Chemistry Consortium, which is funded by EPSRC (EP/F067496), this work made use of the facilities of HECToR, the UK's national high-performance computing service, which is provided by UoE HPCx Ltd at the University of Edinburgh, Cray Inc and NAG Ltd, and funded by the Office of Science and Technology through EPSRC's High End Computing Programme. RGC's visit to JNCASR was supported by the Royal Society, UK. KCS and TSF acknowledge the National Science Foundation (NSF), the Indo-US Science and Technology Forum, and Purdue University for supporting their visits to JNCASR. KCS thanks NSF for financial support via a graduate research fellowship.

**Table 1.**

Vacancy formation energies (VFE) and predicted fractions of vacancies at the phase transition point for each vacancy species in the hydride. $V_{nH}(m)$ represents a cluster of *n* vacancies, while m identifies the configuration.

| Species | VFE* (eV/vacancy) | Fraction of vacancies at transition pressure | | |
|---|---|---|---|---|
| | | 600 K | 700 K | 800 K |
| $V_H$ | 1.41 | $3\times10^{-8}$ | $5\times10^{-7}$ | $4\times10^{-6}$ |
| $V_{2H}(I)$ | 1.04 | $7\times10^{-10}$ | $3\times10^{-8}$ | $4\times10^{-7}$ |
| $V_{2H}(II)$ | 1.13 | $7\times10^{-11}$ | $5\times10^{-9}$ | $9\times10^{-8}$ |
| $V_{3H}(I)$ | 1.07 | $1\times10^{-14}$ | $5\times10^{-12}$ | $2\times10^{-10}$ |
| $V_{3H}(II)$ | 1.10 | $2\times10^{-15}$ | $1\times10^{-12}$ | $8\times10^{-11}$ |
| $V_{3H}(III)$ | 1.10 | $5\times10^{-16}$ | $3\times10^{-13}$ | $2\times10^{-11}$ |

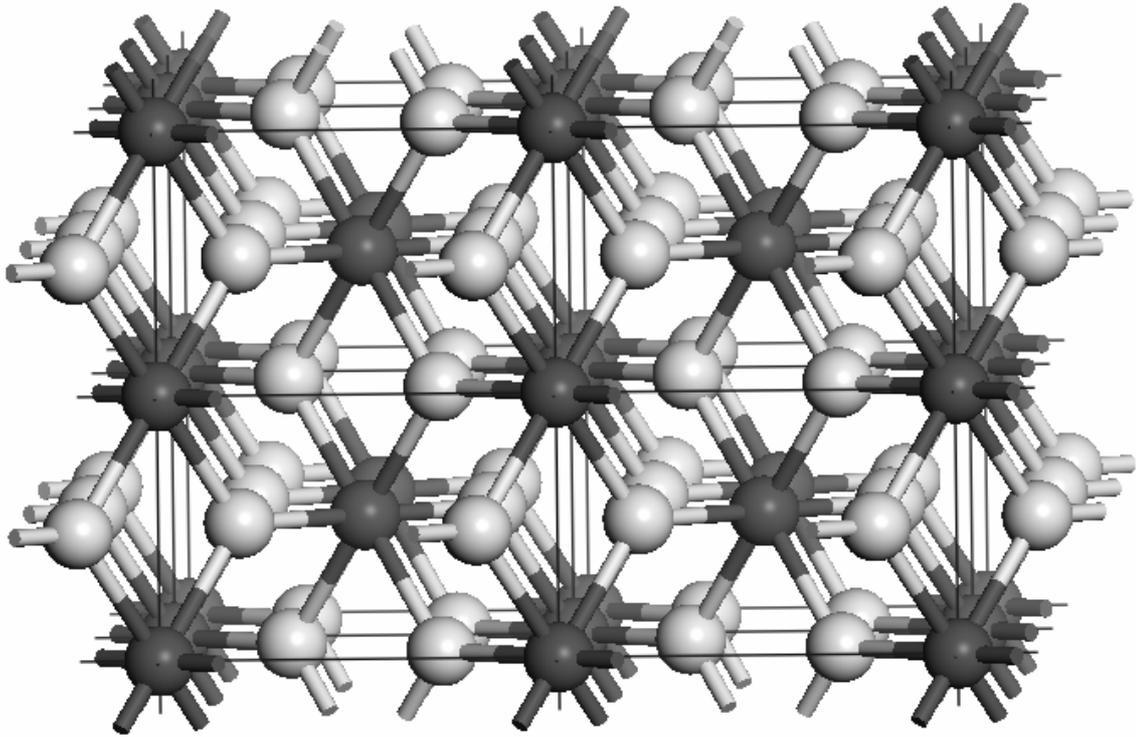

**Fig 1.** The rutile structure of magnesium hydride, shown as a 2×2×2 supercell ($Mg_{16}H_{32-n}$). Dark and light balls represent Mg ions and H atoms, respectively.

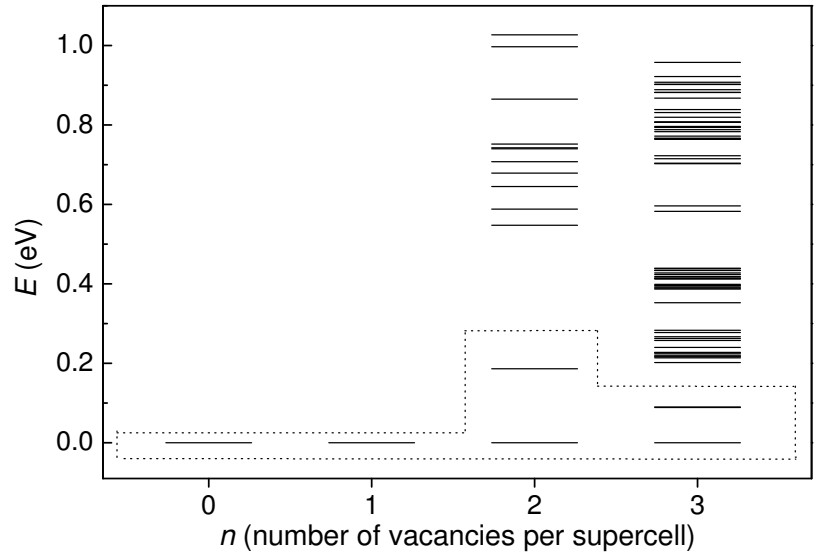

**Fig 2.** Configurational energies (relative to the lowest energy at each composition) in the 2x2x2 supercell Mg$_{16}$H$_{32-n}$. The dotted lines enclose the configurations for which a separate statistical treatment, including vibrational contributions, was performed.

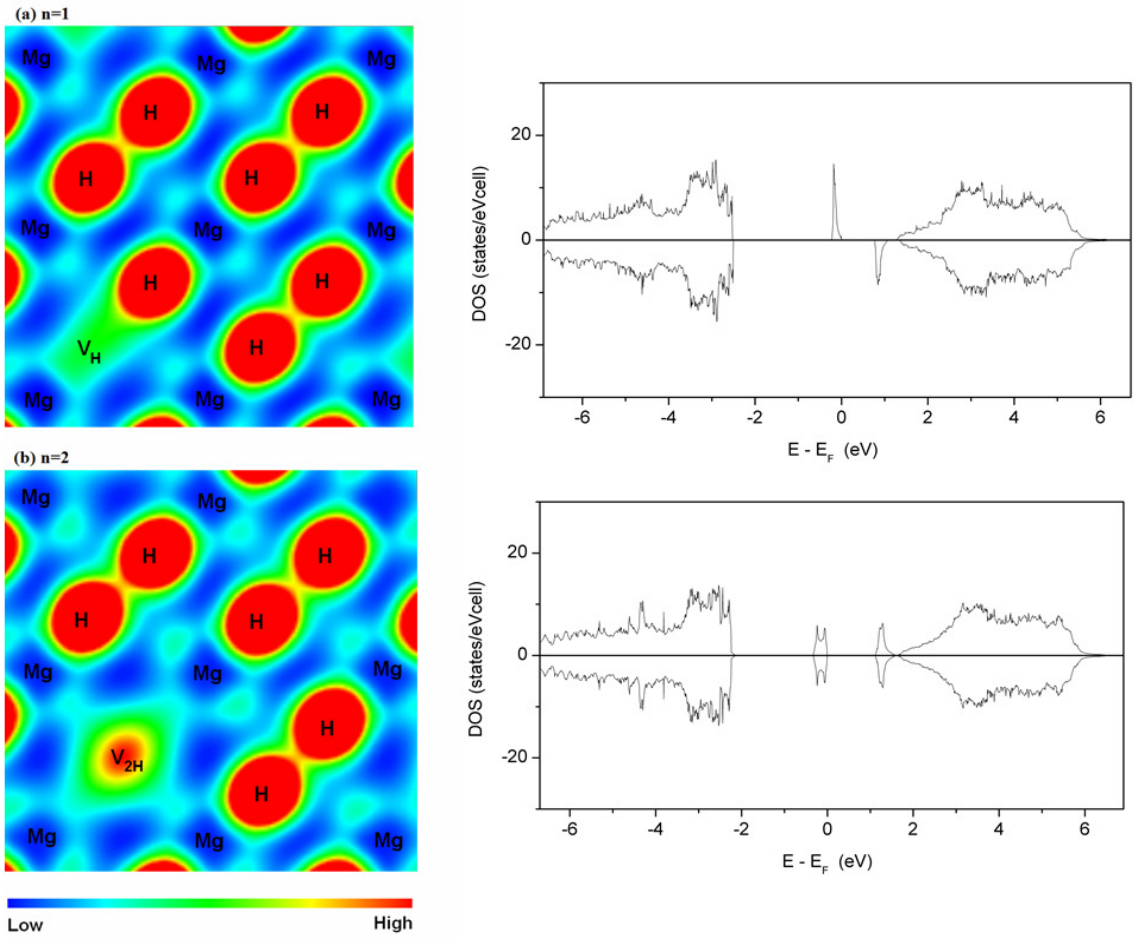

**Fig 3.** (Color online) Charge density distribution in the (001) plane (left) and electronic density of states (right) for a MgH$_2$ cell with (a) an isolated vacancy, (b) a di-vacancy cluster.

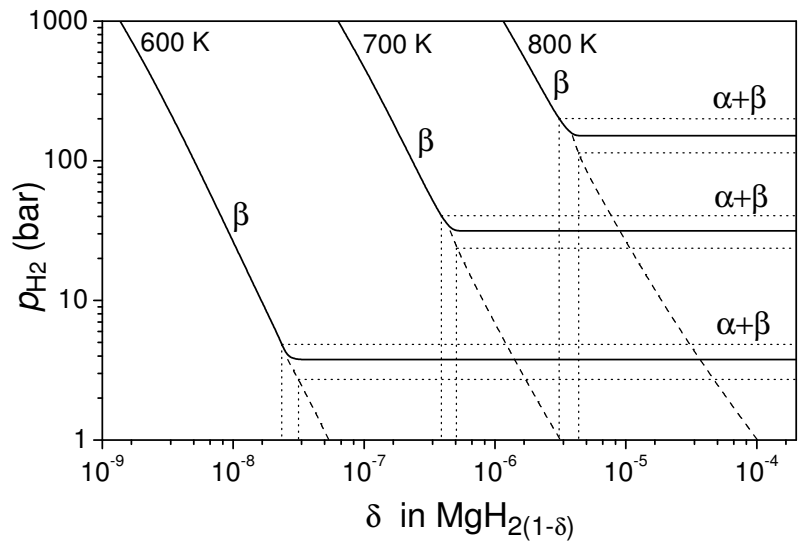

**Fig 4.** Pressure-composition isotherms at pressures above the transformation point. The dashed lines represent the hypothetical extensions of the isotherms in the absence of a phase transition, while the dotted lines represent the error margins of the experimental determination of the transition pressure [32] and their propagation to the theoretically predicted concentrations.

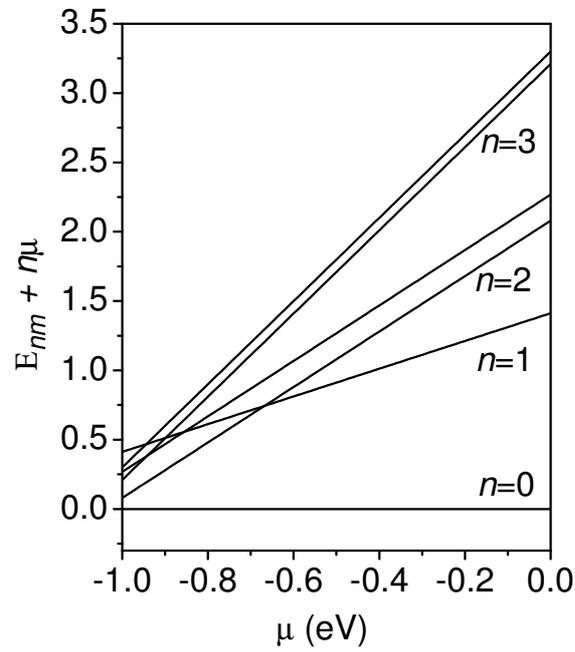

**Fig 5.** Variation in the stability of the most stable configurations for each composition with the hydrogen chemical potential (taken with respect to one half of the energy of the hydrogen molecule).